\documentclass[a4paper,12pt]{article}
\usepackage[utf8]{in
    putenc}
\usepackage{cancel}
\usepackage{ulem}
\usepackage{amsfonts}
\usepackage{amssymb}
\usepackage{graphicx}
\usepackage{amsmath}
\usepackage{enumerate}
\usepackage{mathtools}
\usepackage{subfig}
\usepackage{color}
\usepackage{float}
\usepackage{here}
\usepackage{cite}
\usepackage{mathrsfs}
\usepackage{float,epsfig}
\usepackage{dcolumn}
\usepackage{multirow}
\usepackage{float}
\usepackage{graphicx}
\usepackage{bm}
\usepackage{amsmath,amssymb,amsthm}
\usepackage[colorlinks=true,linkcolor=blue,citecolor=red]{hyperref}
\title{ {\bf\ Anti-de-Sitter-Maxwell-Yang-Mills
   black holes thermodynamics  from nonlocal observables point of view}}
\author{\  H. El Moumni$^{1,2}$\footnote{hasan.elmoumni@edu.uca.ma} \\
\\ 
 {\small $^{1}$ EPTHE, Physics Department, Faculty of Sciences,  Ibn Zohr University, Agadir, Morocco. }
        \\
    {\small $^{2}$ High Energy Physics and Astrophysics Laboratory, Faculty of Science Semlalia,
    }\\
    {\small Cadi Ayyad University, 40000 Marrakesh, Morocco.
    }
}
\date{}
\begin{document} \maketitle

\begin{abstract}

In this paper we analyze the thermodynamic properties
 of the Anti de Sitter black hole in the Einstein-Maxwell-Yang-Mills-AdS gravity (EMYM)  via many approaches and in different thermodynamical ensembles (canonical/ grand canonical). First, we give a concise overview of this phase structure in the entropy-thermal diagram for fixed charges then we investigate this thermodynamical structure in fixed potentials ensemble. The Next relevant step is recalling the nonlocal observables such as holographic entanglement entropy and two-point correlation function to show that both observables exhibit a Van der Waals-like behavior in our numerical accuracy and just near the critical line as the case of the thermal entropy for fixed charges by checking Maxwell's equal area law and the critical exponent. 

In the light of the grand canonical ensemble, we also find a newly phase structure for such a black hole where the critical behavior disappears in the thermal picture as well as in the holographic one.

 \end{abstract}
\newpage
\tableofcontents

\section{Introduction}
Over the last years,  a great emphasis has been put on the application of the Anti-de-Sitter/Conformal field theory correspondence  \cite{1,2} 
which plays a pivotal role in recent developments 
of many physical themes\cite{Hashimoto:1999ut,Faizal:2014dua,3}, in this particular context the thermodynamic of anti-de-Sitter black holes become more attractive for investigation \cite{catastroph}.

In general, black hole thermodynamics has emerged as a fascinating laboratory for testing the predictions of candidate theories of quantum gravity. It has been figured that black holes are associated thermodynamically with a 
entropy and a  temperature\cite{hawking} and a pressure\cite{Kastor}. This association has led to a rich structure of phase picture and a remarkable critical behavior similar to van der Waals liquid/gas phase transition.
\cite{KM,our,our1,our2,ourlovelock,our3,our4,our5,our6,our7,chaos,ref1,ref11,ref2,ref3,ref4,ref5}. Another confirmation of this similarity appear when we employ 
  non-local  observables  such  as  entanglement  entropy, Wilson loop, two point correlation function  and the complexity growth rate \cite{ch12,ch13,ch14,holo,holo1,holo2,holo3,holo4,XXX,ch15,X2,X3,plbmoi,Ghaffarnejad:2018prc}. Meanwhile, these tools are  used  extensively  in quantum information and  to
characterize  phases and thermodynamical behavior
\cite{ee1,ee2,ee4,ee5,ee6,ee7,ee8,Dey:2015ytd,Dudal:2018ztm,anaNPB}.

The black hole charge finds a deep interpretation in
the context of the AdS/CFT correspondence linked to condensed matter physics, the charged black hole introduces a charge density/chemical potential and temperature in the quantum field theory defined on the boundary \cite{cmbase}. In this background,  the charged black hole can be viewed as un uncondensed unstable phase which develops a scalar hair at low temperature and breaks $U(1)$ symmetry near the black hole horizon reminiscing the second order phase transition between conductor and superconductor phases \cite{cm6}, this situation is called the  "s-wave" holographic superconductor.
It has also been shown that "p-wave" holographic superconductor corresponds to vector hair models \cite{cm12,cm14}. The simplest example of p-wave holographic superconductors may be provided by an Einstein-Yang-Mills theory with $SU(2)$ gauge group and no scalar fields, where the electromagnetic gauge symmetry is identified with an $U(1)$ subgroup of $SU(2)$. The other components of the $SU(2)$ gauge field play the role of charged fields dual to some vector operators whose break the $U(1)$ symmetry, leading to a phase transition in the dual field theory.

Motived by all the ideas described above, although the Yang-Mills fields are confined to acting inside nuclei while the Maxwell field dominates outside, the consideration of such theory where the two kinds of field live is encouraged by the existence of exotic and highly dense matter in our universe. In this work, we try to contribute to this rich area by revisiting the phase transition of  Anti-de-Sitter black holes in  Einstein-Maxwell-Yang-Mills (EMYM) gravity.
More especially, we investigate the first and second order phase transition by different approaches including the holographic one and in different canonical ensembles.

This work is organized as follow:  First, we present some thermodynamic properties and phase structure of the EMYM-AdS black holes in (Temperature, entropy)-plane in canonical and grand canonical ensemble. Next, we show in section 3 that the holographic approach exhibit the same behavior,  in other words we recall  the entanglement entropy and two-point correlation function to check the Maxwell's equal area law and   calculating  the critical exponent of the specific heat capacity which is consistent with that of the mean field theory of the Van der Waals in the canonical ensemble near the critical point.  In the grand canonical one, a new phase structure arise 
 where the critical behavior disappears in the thermal as well as the holographic framework.
  The last section is devoted to a  conclusion.

\section{Critical behavior of Einstein-Maxwell-Yang-Mills-AdS black holes in thermal picture}
\subsection{Canonical ensemble}

We start this section by writing the 
 $N$-dimensional  for Einstein-Maxwell-Yang-Mills gravity with a cosmological
constant $\Lambda$  described by the following action  \cite{HabibMazharimousavi:2008ry,Zhang:2014eap}
 \begin{equation}
{\cal I}=\frac{1}{2}\int_{\mathcal{M}} dx^N \sqrt{-g}\left(R-\frac{(N-1)(N-2)}{3}\Lambda-\mathcal{F}_M-\mathcal{F}_{YM}
\right),\label{action}
\end{equation}
where  $R$ is the Ricci scalar while $\Lambda$ is the cosmological constant. Also
$\mathcal{F}_{M}=F_{\mu\nu}F^{\mu\nu}$ and  $\mathcal{F}_{YM}=\textbf{Tr}(F^{(a)}_{\mu\nu}F^{(a)\mu\nu})$ are the  Maxwell invariant and the Yang-Mills invariant respectively, the trace element  stands for  $\textbf{Tr}(.)=\sum_{a=1}^{(N-1)(N-2)/2}(.)$. Varying the action \eqref{action}   with respect to the metric tensor $g_{\mu\nu}$, the Faraday tensor $F_{\mu\nu}$, and the YM tensor $F^{(a)}_{\mu\nu}$, one can obtain the following field equations

\begin{equation}\label{einstein}
G_{\mu\nu}+\Lambda g_{\mu\nu}=T^M_{\mu\nu}+T^{YM}_{\mu\nu}
\end{equation}
\begin{equation}\label{Mmax}
\nabla F^{\mu\nu}=0
\end{equation}
\begin{equation}\label{YMm}
\hat{D}_\mu F^{(a)}_{\mu\nu}=\nabla_\mu F^{(a)\ \mu\nu}+\frac{1}{\Theta}\varepsilon^{(a)}_{(b)(c)} A^{(b)}_\mu F^{(c)\ \mu\nu}=0
\end{equation}
where $G_{\mu\nu}$ is the Einstein tensor, the quantity $\varepsilon^{(a)}_{(b)(c)}$'s stands for the  structure constants of the $\frac{(N-1)(N-2)}{2}$-parameters Lie group $G$, $\Theta$ is coupling constant and 
 $A^{(a)}_\mu$ denote  the $SO(N-1)$ gauge groupe YM potential. We also  note that the internal indices $\{a, b, c, . . .\}$ do not
differ whether in covariant or contravariant form. In addition,
$T^M_{\mu\nu}$ and T$^{YM}_{\mu\nu}$
 are the energy momentum tensor of Maxwell and YM fields with the following formula
 \begin{eqnarray}
 T^M_{\mu\nu}&=&-\frac{1}{2}g_{\mu\nu} F_{\rho\sigma}F^{\rho\sigma}+2F_{\mu\lambda}F^{\lambda}_{\nu}\\
 T^{YM}_{\mu\nu}&=&  \sum_{a=1}^{(N-1)(N-2)/2} \left[ -\frac{1}{2}g_{\mu\nu} F^{(a)}_{\rho\sigma}F^{(a)\rho\sigma}+2F^{(a)}_{\mu\lambda}F^{(a)\lambda}_{\nu}\label{ym10}\right]
 \end{eqnarray}

 \begin{eqnarray}
  &&F_{\mu\nu}^{(a)}=\partial_{\mu}A_{\nu}^{(a)}-\partial_{\nu}A_{\mu}^{(a)}+\frac{1}{2\Theta}
  \varepsilon_{(b)(c)}^{(a)}A_{\mu}^{b}A_{\nu}^{c}, \\
  &&F_{\mu\nu}=\partial_{\mu}A_{\nu}-\partial_{\nu}A_{\mu}.
\end{eqnarray}
Where  $A_{\mu}$ is the usual Maxwell potential.
The metric for such  
  $N$ dimensional spherical black hole may be chosen to  be \cite{Zhang:2014eap}
\begin{equation}\label{frrrrr}
  ds^2=-f(r)dt^2+\frac{1}{f(r)}dr^2+r^2d\Omega_n^2
\end{equation}
in which,    $d\Omega_n^2$  represents the  volume of the unit $n$-sphere 
which can be expressed in the standard
spherical form%
\begin{equation}
d\Omega _{N-2}^{2}=d\theta _{1}^{2}+\sum_{i=2}^{N-3}
\Pi_{j=1}^{i-1}\sin ^{2}\theta _{j}\;d\theta _{i}^{2} \ \ \text{ where }\ \ \ \  0\leq \theta _{1}\leq \pi ,0\leq \theta _{i}\leq 2\pi .
\end{equation}%

In order to find the electromagnetic field, we recall the following radial gauge potential ansatz $A_\mu=h(r)\delta_\mu^0$ which obeys to  the Maxwell field equations \eqref{Mmax} with the following solution
\begin{equation}
\frac{dh(r)}{dr}=\frac{c}{r^{N-2}}
\end{equation}
where $c$ is an integration constant related to electric charge  $C$ of the solutions.
To solve the YM field, Eq.\eqref{YMm}, we use the magnetic Wu–Yang ansatz of the gauge potential \cite{YM12,YM16} given by 

\begin{eqnarray}
A^{(a)} &=&\frac{Q}{r^{2}}\left( x_{i}dx_{j}-x_{j}dx_{i}\right) \\
2 &\leq &i\leq N-1,  \notag \\
1 &\leq &j\leq i-1  \notag \\
1 &\leq &\left( a\right) \leq \left( N-1\right) (N-2)/2  \notag
\end{eqnarray}%
where we imply (to have a systematic process) that the super indices $a$ is
chosen according to the values of $i$ and $j$ in order. For instance, we
present some of them 

\begin{equation}
\begin{array}{c}
A^{\left( 1\right) }=\frac{Q}{r^{2}}\left( x_{2}dx_{1}-x_{1}dx_{2}\right) \\ 
A^{\left( 2\right) }=\frac{Q}{r^{2}}\left( x_{3}dx_{1}-x_{1}dx_{3}\right) \\ 
A^{\left( 3\right) }=\frac{Q}{r^{2}}\left( x_{3}dx_{2}-x_{2}dx_{3}\right) \\ 
A^{\left( 4\right) }=\frac{Q}{r^{2}}\left( x_{4}dx_{1}-x_{1}dx_{4}\right) \\ 
A^{\left( 5\right) }=\frac{Q}{r^{2}}\left( x_{4}dx_{2}-x_{2}dx_{4}\right) \\ 
A^{\left( 6\right) }=\frac{Q}{r^{2}}\left( x_{4}dx_{3}-x_{3}dx_{4}\right) \\ 
A^{\left( 7\right) }=\frac{Q}{r^{2}}\left( x_{5}dx_{1}-x_{1}dx_{5}\right) \\ 
A^{\left( 8\right) }=\frac{Q}{r^{2}}\left( x_{5}dx_{2}-x_{2}dx_{5}\right) \\ 
A^{\left( 9\right) }=\frac{Q}{r^{2}}\left( x_{5}dx_{3}-x_{3}dx_{5}\right) \\ 
A^{\left( 10\right) }=\frac{Q}{r^{2}}\left( x_{5}dx_{4}-x_{4}dx_{5}\right)
\\ 
...%
\end{array}%
\end{equation}%
in which $r^{2}=\overset{N-1}{\underset{i=1}{\sum }}x_{i}^{2}.$
The YM field 2-forms are defined by the expression%
\begin{equation}
F^{\left( a\right) }=dA^{\left( a\right) }+\frac{1}{2Q}\varepsilon_{\left( b\right)
\left( c\right) }^{\left( a\right) }A^{\left( b\right) }\wedge A^{\left(
c\right) }.
\end{equation}


In general for $N$ we must have $\left( N-1\right) \left( N-2\right) /2$
gauge potentials. The integrability conditions 
\begin{equation}
dF^{\left( a\right) }+\frac{1}{Q}\varepsilon_{\left( b\right) \left( c\right)
}^{\left( a\right) }A^{\left( b\right) }\wedge F^{\left( c\right) }=0
\end{equation}%
are easily satisfied by using (28). The YM equations 
\begin{equation}
d\ast F^{\left( a\right) }+\frac{1}{Q}\varepsilon_{\left( b\right) \left( c\right)
}^{\left( a\right) }A^{\left( b\right) }\wedge \ast F^{\left( c\right) }=0
\end{equation}%
also are all satisfied. The energy-momentum tensor (4), becomes after%
\begin{equation}
\sum_{a=1}^{\left( N-1\right) \left( N-2\right) /2}%
\left[ F_{\lambda \sigma }^{\left( a\right) }F^{\left( a\right) \lambda
\sigma }\right] =\frac{\left( N-3\right) \left( N-2\right) Q^{2}}{r^{4}}
\end{equation}%
with the non-zero components%
\begin{eqnarray}
T_{00} &=&\frac{\left( N-3\right) \left( N-2\right) Q^{2}f\left( r\right) }{%
2r^{4}} \\
T_{11} &=&-\frac{\left( N-3\right) \left( N-2\right) Q^{2}}{2r^{4}f\left(
r\right) }  \notag \\
T_{22} &=&-\frac{\left( N-3\right) \left( N-6\right) Q^{2}}{2r^{2}}  \notag
\\
T_{AA} &=&-\frac{\left( N-3\right) \left( N-6\right) Q^{2}}{2r^{2}}\Pi_{i=1}^{%
A-2}\sin ^{2}\theta _{i}  \notag \\
2 &<&A\leq N-1.  \notag
\end{eqnarray}

Using the Eq.\eqref{einstein} and after some simplifications, one can find
that   the metric function $f(r)$ has the following form 
is given by
\begin{eqnarray}
f(r)&=&1-\frac{2m}{r^{n-1}}-\frac{\Lambda}{3}r^2+\frac{2(n-1)C^2}{nr^{2n-2}}-\frac{(n-1) Q^2}{(n-3) r^2},
\label{fr}
\end{eqnarray}
Where $N=n+2$, one can note in the  particular case for $n=3$ the last term of Eq.\eqref{fr} diverges, involving  an unusual logarithmic term in Yang-Mills charge \cite{HabibMazharimousavi:2008ry}.
For this gravity background the parameter $m$ is related to the mass of such black hole, while $C$ and $Q$ are the charges of Maxwell field and Yang-Mills field respectively.
Following previous literature \cite{Kastor,KM}, one can finds a close connection between the cosmological constant and pressure as 
$P=-\frac{\Lambda}{8\pi}
$, leading to the following expressions of 
 Hawking temperature, mass and entropy of such black hole in term of the horizon radius $r_+$ \begin{eqnarray}
  &&T=\frac{f'(r_+)}{4\pi}=\frac{n-1}{4\pi r_+}+\frac{2(n+1)P}{3}r_+ -\frac{(n-1) Q^2}{4 \pi  r^3}
  -\frac{(n-1)^2C^2}{2\pi n r_+^{2n-1}}, \label{equ:t}\\
  &&M=\frac{n\omega_n}{48\pi}\big(8\pi P r_+^{n+1}+\frac{3 (n-1) Q^2 r^{n-3}}{3-n}+3r_+^{n-1}
  +\frac{6(n-1)C^2}{nr_+^{n-1}} \big),\label{equ:M}\\
  && S=\frac{\omega_n r_+^n}{4} \label{equ:s},
\end{eqnarray}

 The Yang-Mills potential $\Phi_Q$ and the electromagnetic one $\Phi_C$ can be written as
\begin{eqnarray}\label{potential1}
  &&\Phi_Q=\left(\frac{\partial M}{\partial Q}\right)_{C,P,r_+}=\frac{\omega_n (n-1)nQ}{8\pi(3-n)}r_+^{n-3}, \\ \label{potential2}
  &&\Phi_C=\left(\frac{\partial M}{\partial C}\right)_{Q,P,r_+}=\frac{\omega_n (n-1)C}{4\pi r_+^{n-1}},
\end{eqnarray}
where $\omega_n=\frac{2\pi^{(n+1)/2}}{\Gamma(\frac{n+1}{2})}$ is the volume of the unit $n$-sphere.
In fact,  according to the interpretation of the black hole mass $M$ as an enthalpy \cite{Kastor} in the extended phase space context, the free energy $\mathcal{F}$ of black hole can be written as
\begin{eqnarray}
  \mathcal{F}=M-T\cdot S
\label{F}
\end{eqnarray}
and the heat capacity is given by
\begin{equation}
C_Q=\left. T \frac{\partial S}{\partial T}\right|_Q
\end{equation}

It straightforward to show that obtained  quantities   Eq.\eqref{equ:t}, Eq.\eqref{equ:M} and Eq.\eqref{equ:s},  obey  to the first law of black
hole thermodynamics in the extended phase space
\begin{equation}
    dM=T  dS+\Phi_Q  dQ+\Phi_C  dC+V  dP,
\end{equation}
where $V$ is the Legendre transform of the pressure, which
denotes the thermodynamic volume with 
\begin{equation}
V=\left(\frac{\partial M}{\partial P}\right)_{S,\Phi_Q,\Phi_C}=\frac{n\omega_n \pi}{3}r^{n+1}.
\end{equation}
In addition to this, using scaling argument, 
the corresponding Smarr formula is\begin{equation}
  M=\frac{n}{n-1}TS+\Phi_C C+\frac{1}{(n-1)}\Phi_Q Q-\frac{2}{n-1}VP.
\end{equation}
 Without loss of generality, 
 inserting Eq.\eqref{equ:s} into Eq.\eqref{equ:t}, we can get the entropy  Hawking temperature relation of such black hole, namely
 \begin{eqnarray}\label{ts}\nonumber
T&=&\frac{1}{12 \pi  n}\left[6 C ( n-1)^2 \left(4^{\frac{1}{n}} \left(\frac{S}{\omega_n}\right)^{\frac{1}{n}}\right)^{1-2 n}+\pi  2^{\frac{2}{n}+3} n (n+1) P
   \left(\frac{S}{\omega_n}\right)^{\frac{1}{n}}-3 (n-1) n Q^2
   4^{\frac{-3}{n}} \left(\frac{S}{\omega_n}\right)^{\frac{-3}{n}}\right.\\
   &+&\left.3\ 4^{-1/n} (n-1) n \left(\frac{S}{\omega_n}\right)^{-1/n}\right]
\end{eqnarray}

\begin{center}
\begin{figure}[!ht]
\begin{tabbing}
\hspace{9cm}\=\kill
\includegraphics[scale=1]{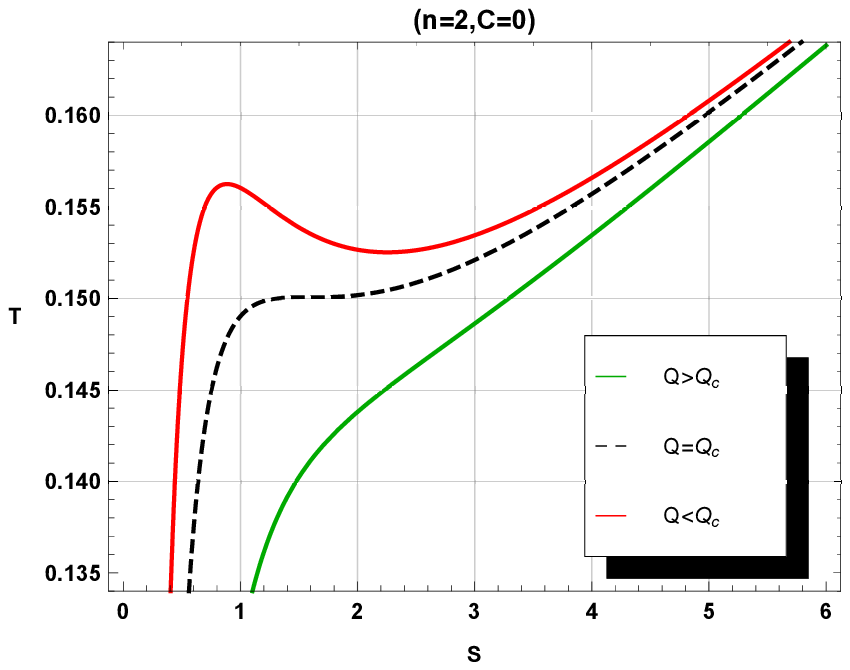}\> \includegraphics[scale=1]{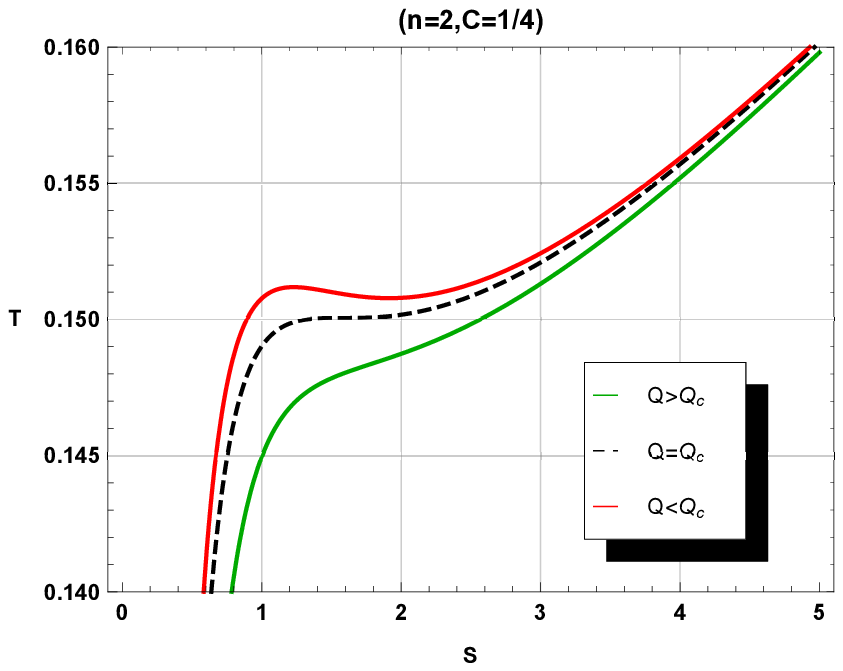} 
\end{tabbing}
\vspace*{-.2cm} \caption{\footnotesize The temperature as a function of the entropy for different  values of charge $C$. {\bf(Left)}   $C=0$. {\bf (Right)}   $C\neq 0$.}\label{fig1}
\end{figure}
\end{center}

This relation is depicted  in Fig.\ref{fig1}, indeed it has been shown that there is a Van der Waals-like phase transition, furthermore a direct   confirmation comes from the solution of the following system

\begin{equation}
\left(\frac{\partial T}{\partial S} \right)_{Q}= \left(\frac{\partial^2 T}{\partial S^2} \right)_{Q} =0.
\end{equation}

which reveal the existence of a critical point. The critical charge, entropy, and temperature are given in the following Tab.\ref{tabcritic} with for all the rest of the paper we keep $n=2$.

\begin{table}[h!t]
\begin{center}
\begin{tabular}[t]{l|l|l|l|l|}\cline{2-4}
  &$Q_c$ & $S_c$ & $T_c$\\\hline\hline
${\tt C=0}$& $\frac{1}{2\sqrt{3}}$	  & $\frac{\pi}{2}$  & $\frac{\sqrt{2}}{3\pi}$ 
\\\hline
${\tt C\neq 0}$&	$\frac{1}{6} \sqrt{3-36 C^2}$  &  $\frac{\pi}{2}$ & $\frac{\sqrt{2}}{3\pi}$ 
\\\hline
\hline
\end{tabular}
\end{center}
\caption{\footnotesize Coordinates of the critical points for different values of $C$ in $(T,S)$-diagram.}\label{tabcritic}
\end{table}

\begin{center}
\begin{figure}[!ht]
\begin{tabbing}
\hspace{9cm}\=\kill
\includegraphics[scale=.9]{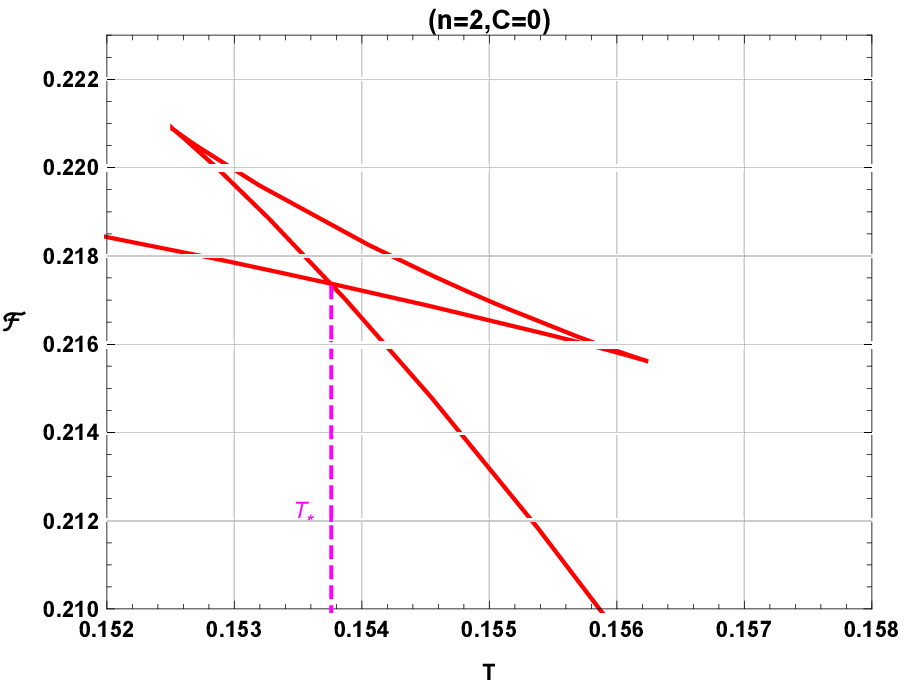}\> \includegraphics[scale=.9]{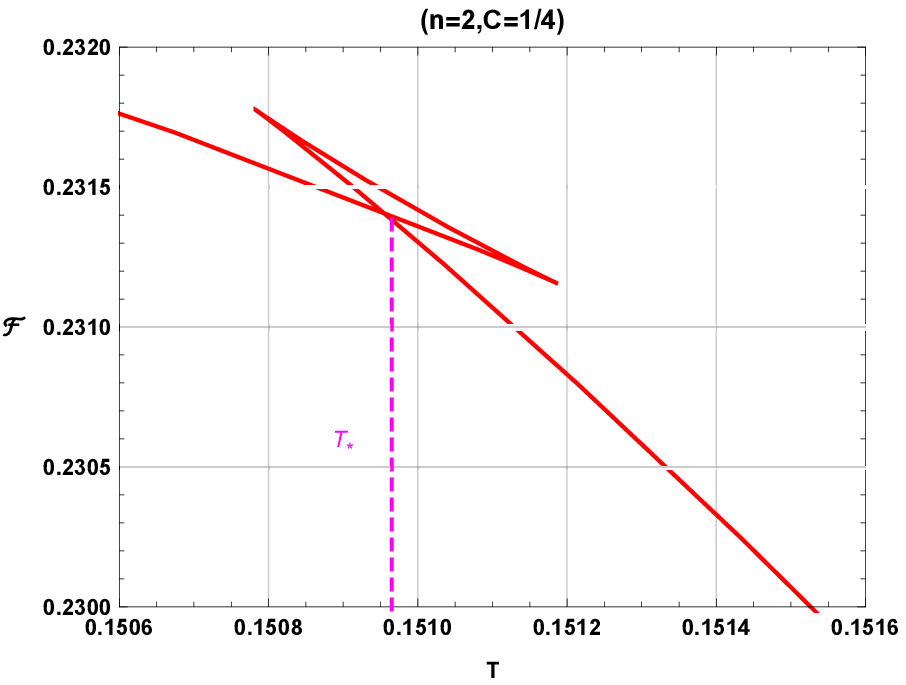} 
\end{tabbing}
\vspace*{-.2cm} \caption{\footnotesize The Helmholtz free energy in function of the entropy for EMYM-AdS black holes for different values of the charge $C$.
 {\bf(Left)}   $C=0$. {\bf (Right)}   $C\neq 0$.}\label{fig2}
\end{figure}
\end{center}

An important remark can be observed here is that both quantities $T_c$ and $S_c$ are insensible the charge $C$.
The behavior of the free energy  $\mathcal{F}$ with respect to the temperature may be investigated by plotting in the Fig.\ref{fig2}  the graph $\mathcal{F}-T$ for a fixed value of charge $Q$ under the critical one. 
 From  this plot, we can observe the characteristic swallow-tail 
which guarantees the existence of the Van der Waals-like phase transition.

Using  the Fig.\ref{fig2}, we  can derive numerically  the coexistence temperature    $T_{\star}$ needed  in the Maxwell's equal area law construction 
\begin{equation}\label{euqalarea}
A_1\equiv \int_{S_1}^{S_2} T(S) dS-T_\star(S_2-S_1)=T_\star(S_3-S_2)-\int_{S_2}^{S_3} T(S) dS\equiv A_2,
\end{equation}
where $S_3$, $S_2$ and   $S_1$  are the solutions  of 
$T(S)=T_\star$ in descending order, in addition to the numerical values  of these point and the we report in in Tab.\ref{tab2} the 
 area  $A_1$ and $A_2$ in Maxwell's law Eq.\eqref{euqalarea}

\begin{table}[h!t]
\begin{center}\begin{tabular}{|l|l|l|l|l|l|l|}\cline{2-5}
 \hline
        C       &$T_\star$ &   $S_1$ &   $S_2$&    $S_3$    & $A_1$   &  $A_2$    \\ \hline\hline
     ${\tt 0}$  & 0.15376 & 0.638237 & 1.52994 &3.13241& 0.383499& 0.38350
        \\ \hline\hline
       
        ${\tt \frac{1}{4}}$ &150965& 1.05043& 11.54616& 2.25932& 00.18248& 0.1825
        \\ \hline

  \end{tabular}
\end{center}
\caption{\footnotesize Check of the equal area law in the $T-S$ plane for different $C$.}\label{tab2}
\end{table}

Obviously, the area $A_1$ equals to area  $A_2$  for different $C$, so the equal area law doesn't break.
For the second phase transition, we know that near the critical point, there is always a  linear relation with slope equal to $3$ \cite{Zhang:2014eap,bbb} 
\begin{equation}\label{ord2}
log |T-T_c|=3\ log|S-S_c|+constant
\end{equation}
in this context,  the heat capacity behaves  like
 \begin{equation}
 C_Q\sim (T-T_c)^{-2/3}
 \end{equation}
where the critical exponent of the second order phase is $-2/3$, which is consistent with the mean field theory.

\subsection{Grand canonical ensemble}
Having described the thermodynamical behavior of the EMYM-AdS black hole with a fixed charge, by showing the occurrence of the first and second phase transition, we will focus on this section to the phase structure when the potentials are kept fix.

To facilitate the calculation of relevant quantities, it is
convenient to reexpress the Hawking temperature  as a function of entropy, Yang-Mills and  electromagnetic potentials,  inserting Eq.\ref{potential1} and Eq.\eqref{potential2} into Eq.\eqref{ts} on can write

\begin{eqnarray}\label{tsgc}
T&=&\frac{1}{3\pi   2^{\frac{2 (n+1)}{n}}}
\left(\frac{S}{\omega_n}\right)^{-1/n}\\ \nonumber
&\times&
\left[\frac{96 \pi ^2 \left(-\frac{2^{\frac{n+8}{n}} (n-3)^2 \Phi_Q^2
   \left(4^{\frac{1}{n}} \left(\frac{S}{\omega_n}\right)^{\frac{1}{n}}\right)^{-2 n}
   \left(\frac{S}{\omega_n}\right)^{4/n}}{n-1}-n \Phi_c^2\right)}{n^2 \omega
   _n^2}-  16^{\frac{1}{n}}  \Lambda(n+1) \left(\frac{S}{\omega
   _n}\right)^{2/n}+3 (n-1)\right]
\end{eqnarray}

In our assumptions, where $n=2$ and $\Lambda=-1$, the  Eq.\eqref{tsgc} reduces to
\begin{equation}\label{tcan}
T=\frac{S-\pi  \left(\Phi _c^2+\Phi _Q^2-1\right)}{4 \pi ^{3/2} \sqrt{S}}
\end{equation}
which is in agreement  with the result of \cite{cano25,cano26} if we set $\Phi_Q=0$. Now we are able to write easily 
\begin{equation}
\left(\frac{\partial T}{\partial S}\right)_{\Phi_Q,\Phi_c}=\frac{-S-3 \pi  \left(\Phi_c^2+\Phi_Q^2-1\right)}{16 \pi ^{3/2} S^{5/2}}
\end{equation}

\begin{equation}\label{div2}
\left(\frac{\partial^2 T}{\partial S^2}\right)_{\Phi_Q,\Phi_c}=\frac{S+\pi  \left(\Phi_c^2+\Phi_Q^2-1\right)}{8 \pi ^{3/2} S^{3/2}}
\end{equation}

\begin{center}
\begin{figure}[!ht]
\begin{tabbing}
\hspace{9cm}\=\kill
\includegraphics[scale=.8]{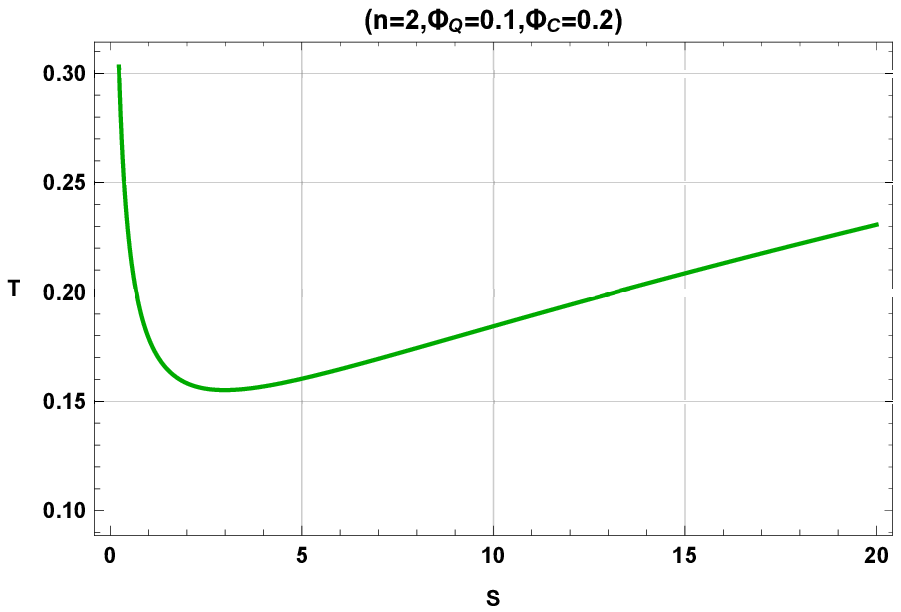}\> \includegraphics[scale=.8]{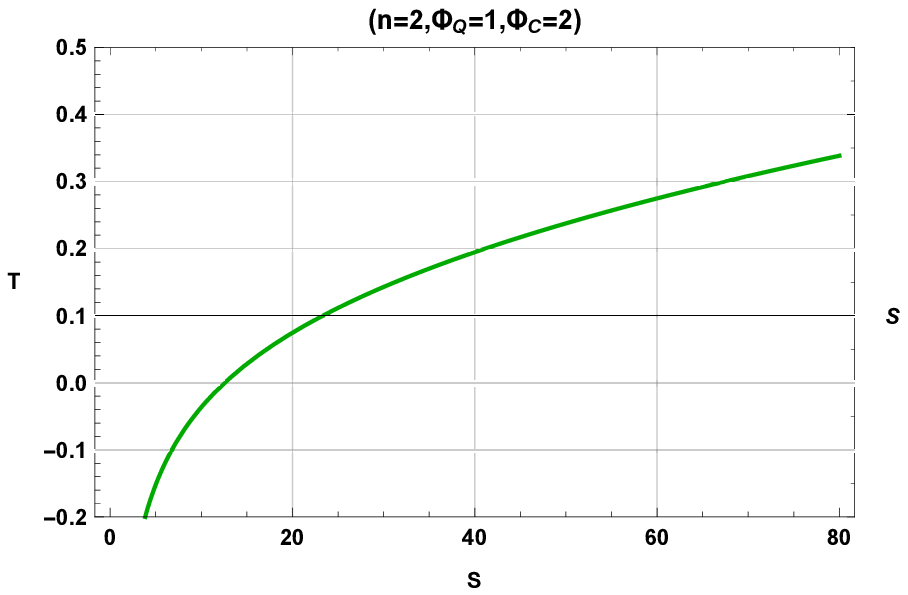} 
\end{tabbing}
\vspace*{-.2cm} \caption{\footnotesize The temperature as a function of the entropy for different potentials. {\bf(Left)} $\Phi_c^2+\Phi_Q^2<1$. {\bf (Right)} $\Phi_c^2+\Phi_Q^2>1$.}\label{figcan}
\end{figure}
\end{center}

The solution of the $\left(\frac{\partial T}{\partial S}\right)_{\Phi_Q,\Phi_c}=0$ can be derived as 
\begin{equation}\label{s1}
S_1= -\pi  \left(\Phi_c^2+\Phi_Q^2-1\right)
\end{equation}
with the condition $\Phi_c^2+\Phi_Q^2<1$ should be verified to ensure that the entropy in Eq.\eqref{s1} is positive. In the other  case $
\Phi_c^2+\Phi_Q^2>1$, one can't find  meaningful root of the  equation  $\left(\frac{\partial T}{\partial S}\right)_{\Phi_Q,\Phi_c}=0$. Substituting Eq.\eqref{s1} into Eq.\eqref{div2} we obtain the following constraint

\begin{equation}
\left.\left(\frac{\partial^2 T}{\partial S^2}\right)_{\Phi_Q,\Phi_c}\right|_{S=S_1}= \frac{1}{8 \pi ^3 \left(-\Phi_c^2-\Phi_Q^2+1\right)^{3/2}}>0
\end{equation}

implying that, no critical point is observed in the $T-S$ diagram. This observation differs from the  result in the previous section where the charges are kept fixed, consolidating the assertion
 that the thermodynamics in the grand canonical ensemble is quite different from that in the canonical one.
In Fig.\ref{figcan}, we depict the Hawking temperature for both the case $\Phi_c^2+\Phi_Q^2<1$  and  $\Phi_c^2+\Phi_Q^2>1$ and  one can see that  there exists minimum temperature for the left panel . Substituting Eq.\eqref{s1} into Eq.\eqref{tcan}, the minimum temperature can be obtained as 
\begin{equation}
T_{min}=\frac{\sqrt{-\Phi_c^2- \Phi_Q^2+1}}{2 \pi }.
\end{equation}
While, the Hawking temperature increases monotonically in the right panel.

Having obtained the phase picture of the thermal entropy of the AdS-Maxwell-Yang-Mills black hole the canonical/grand canonical ensemble, we will now revisit the phase structure of the entanglement entropy and two-point correlation function to see whether they have similar phase structure in each thermodynamical ensemble.

\section{Phase transitions of Einstein-Maxwell-Yang-Mills-AdS black holes in holographic picture }
\subsection{Holographic entanglement entropy}

First, let us provide a concise review of some generalities about the holographic entanglement entropy.
For a given quantum field theory described by a density matrix $\rho$, with  $A$ is some region of a Cauchy surface of spacetime and $A^c$ stands for its complement, the von Neumann entropy which traduce the entanglement between these two regions   
\begin{equation}
S_{A} = -\mathrm{Tr}_{A}{(\rho_{A}\log{\rho_{A}})}\,,
\end{equation}
with $\rho_{A}$ is the reduced density matrix of  given by
$A$  $\rho_{A}=\mathrm{Tr}_{A^{c}}{(\rho)}$.  
  Ryu and Takayanagi  propose a simple geometric way  to evaluate   the entanglement entropy as  \cite{7,8} \begin{equation}\label{288}
S_{A} = \frac{\text{Area}(\Gamma_A)}{4 G_N}\,,
\end{equation}
in which $\Gamma_A$ denote a codimension-2 minimal surface with boundary condition $\partial \Gamma_A=\partial A$, and $G_{N}$ stands for the gravitational Newton's constant. In our black hole model we choose the region $A$ to be
a spherical cap on the boundary delimited by $\theta\leq \theta_0$, and  
the minimal surface can be  parametrized   by the function $r{(\theta)}$.  According to definition of the area and Eq.\eqref{frrrrr} and Eq.\eqref{288} on can show that the holographic entanglement entropy is governed by  

\begin{equation}\label{AA}
S_{A}  = \frac{\omega_{n-2}}{4} \int_{0}^{\theta_{0}} r^{n-2}\sin^{n-2}{\theta}\sqrt{\frac{(r')^{2}}{f{(r)}}+r^{2}} d\theta\,,
\end{equation}
where the notation prime  denotes the derivative with respect to $\theta$ e.g. $r' \equiv \frac{dr}{d\theta}$.
Treating the  Eq.\eqref{AA} as a Lagrangian and  solving the  equation of motion given by 
\begin{eqnarray}\nonumber
& &r'(\theta )^2 [\sin \theta  r(\theta )^2 f'(r)-2 \cos \theta r'(\theta )]-2 r(\theta ) f(r) [r(\theta ) (\sin \theta  r''(\theta ) 
+\cos \theta  r'(\theta ))
-3 \sin \theta  r'(\theta )^2]\\ &+&4 \sin (\theta ) r(\theta )^3 f(r)^2=0.\label{sysEE}
\end{eqnarray}
Due to the difficulty to  find an analytical form  of the solution $r{(\theta)}$, we will perform a numerical calculation with adopting 
 the following boundary conditions 
\begin{equation}\label{bc}
 r'(\theta)=0,\;\; r(0)=r_0.
\end{equation}

\begin{center}
\begin{figure}[!h]
\begin{tabbing}
\hspace{9cm}\=\kill
\includegraphics[scale=.78]{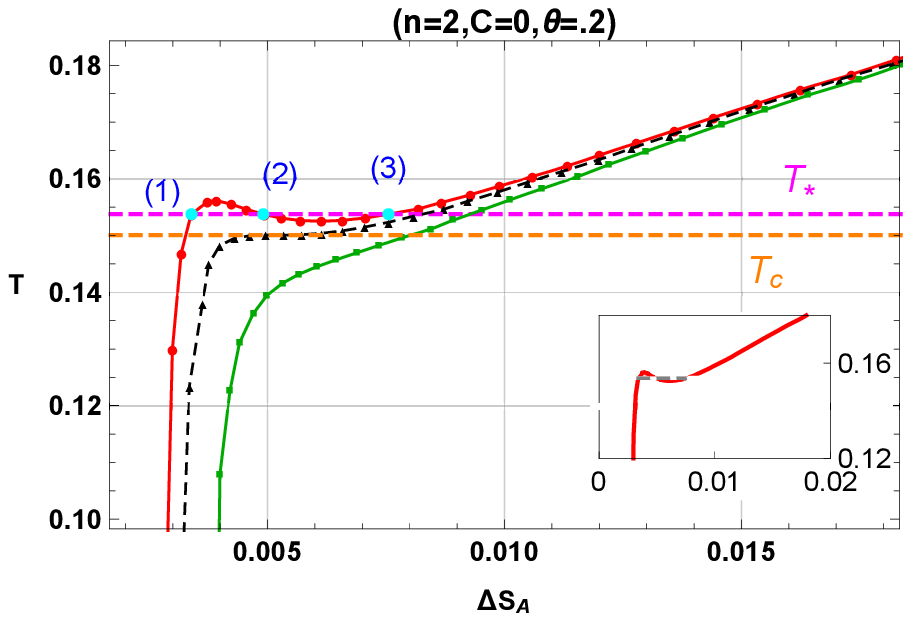}\> \includegraphics[scale=.78]{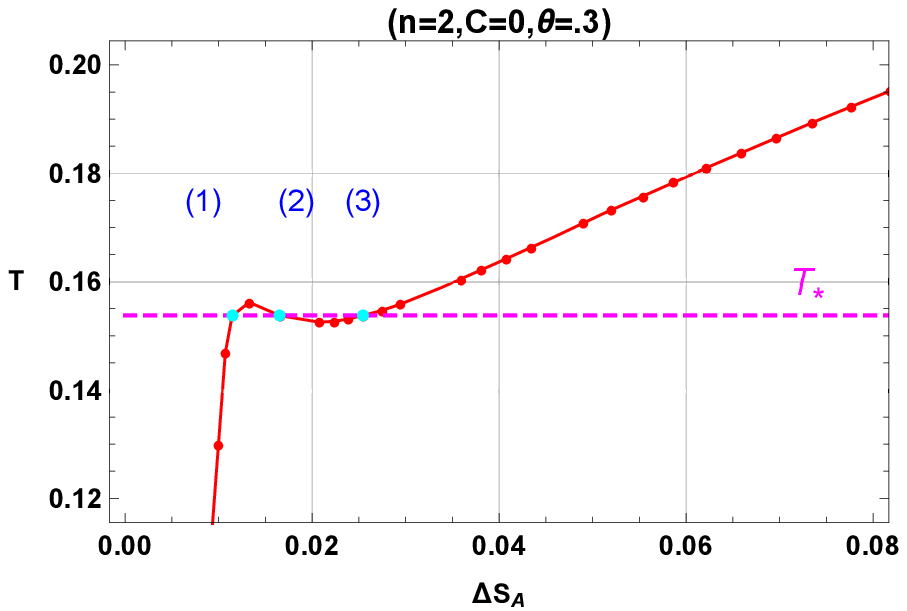} \\
\includegraphics[scale=.798]{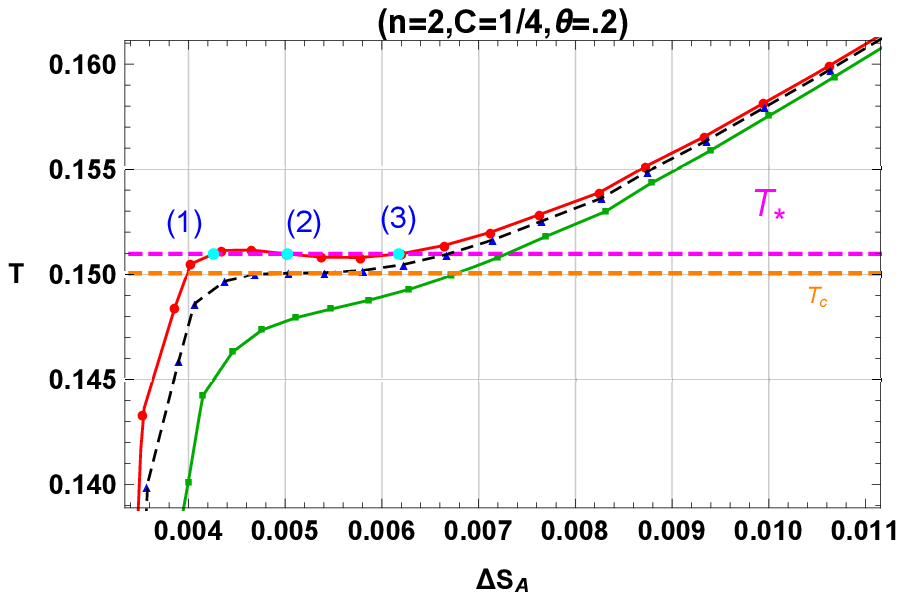}\> \includegraphics[scale=.78]{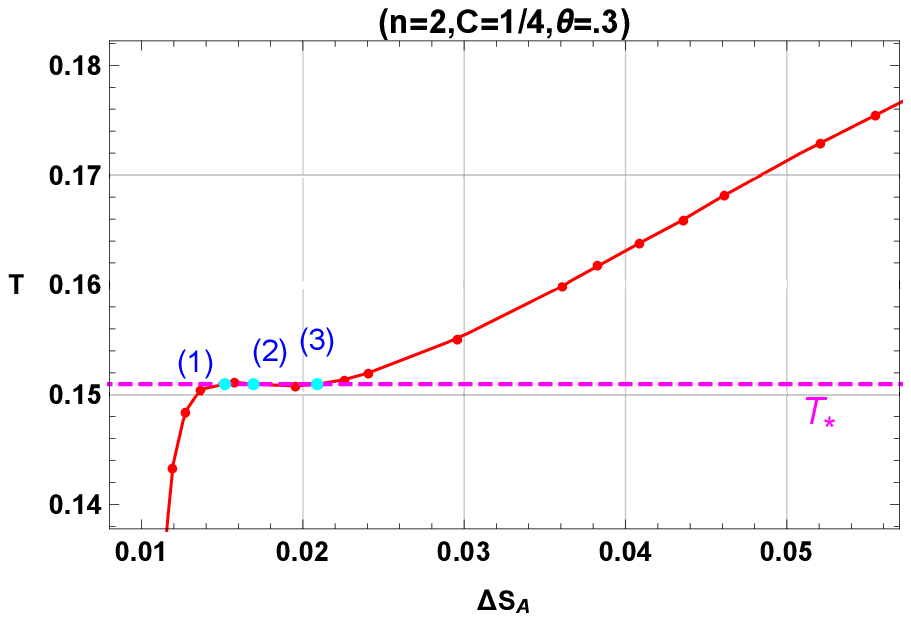} \\
\end{tabbing}
\vspace*{-.2cm} \caption{\footnotesize	Plot of isocharges on the $(T,\Delta S_{A})$-plan, for different values of $\theta_0,\; C$. For all panels:
 the values of the charge are $Q=.9 Q_c$ (red), $Q=Q_{c}$ (dashed black) and $Q= 2 Q_c$ (green).
 The coexistence phase isotherm $T_\star$ (dashed magenta line) is obtained from the free energy  Figure.\ref{fig2} and  the critical temperature (dashed orange line).
 For all curves, we also show the data points which are used to create the interpolation.}\label{fig3}
\end{figure}
\end{center}

Knowing that the entanglement entropy is divergent at the boundary, we regularized it by subtracting   the area of the minimal surface in pure AdS whose boundary is also $\theta=\theta_{0}$
with
\begin{equation}
r_{AdS}{(\theta)} = L\left(\left(\frac{\cos{\theta}}{\cos{\theta_{0}}}\right)^{2}-1\right)^{-1/2}\,.
\end{equation}
We label the regularized entanglement entropy  by $\Delta S_A$ and for our numerical calculation we choose 
 $\theta_{0} = 0.2$, and $ 0.3$  while 
Ultra Violet cutoff is    chosen    to be $\theta_{c} =  0.199$ and  $0.299$ respectively.
To compare with the phase structure in the thermal picture, we will study the relation between the entanglement entropy and the Hawking temperature representing the temperature of the dual field theory, this relation is depicted in  
 Fig.\ref{fig3}  for different values of Maxwell's charge with a fixed Yang-Mills one near the critical point and the chosen $\theta_{0}$. 

For each panel, 
 the red lines are associated with a charge less than the critical one while equal to the critical charge are depicted in black dashed lines and green lines correspond to a charge upper than the critical. Especially, the first order
phase transition temperature $T_\star$ and second-order phase transition temperature $T_c$ are plotted by magenta and orange dashed line respectively. As can be seen in all these plots, the Van der Waals-like phase structure can also be observed in the $T-\Delta S_A$diagram.  
Particularly, the coexistence  temperature $T_\star$ and second-order phase transition temperature $T_c$ are exactly the same as that in the thermal entropy structure.

Adopting the same steps as in the thermal picture, we will also check numerically whether Maxwell’s equal area law holds
\begin{equation}
A_{1} = \int_{\Delta S_{A}^{(1)}}^{\Delta S_{A}^{(2)}} T{(\Delta S_{A},Q)} d\Delta S_{A} - T_{*}(\Delta S_{A}^{(2)}-\Delta S_{A}^{(1)})
\end{equation}
\begin{equation}
A_{2} =  T_{*}(\Delta S_{A}^{(3)}-\Delta S_{A}^{(2)}) - \int_{\Delta S_{A}^{(2)}}^{\Delta S_{A}^{(3)}} T{(\Delta S_{A},Q)} d\Delta S_{A}
\end{equation}

with the quantities  $\Delta S_{A}^{(1)}$, $\Delta S_{A}^{(2)}$ and $\Delta S_{A}^{(3)}$ are 
 roots of the equation $T_{*} = T{(\Delta S_{A},Q)}$ in the ascending order. The Maxwell's equal area law stipulate that 
\begin{equation}
A_{1} = A_{2}.
\end{equation}
We tabulate in Tab.\ref{Table3} the values of  the both areas $A_{1}$ and $A_{2}$ for the chosen  $\theta_{0}$, the  charge $C$ as well as  the relative error between $A_{1}$ and $A_{2}$
taken to be the difference between $A_{1}$ and $A_{2}$ divided by their average.

\begin{center}
\begin{table}[!ht]
\centering\small
\begin{tabular}{|l|l|l|l|l|l|l|l|l|}
\hline
          $C$        &         $\frac{Q}{Q_c}$                   & $\theta_0$ & $\Delta S_{A}^{(1)}$ & $\Delta S_{A}^{(2)}$ & $\Delta S_{A}^{(3)}$ &  $A_{1}$ & $A_{2}$ & Relative error  \\ \hline\hline
          \multirow{4}{*}{$0$} & \multirow{2}{*}{\textbf{0.9}} & 0.2  &  $0.00339417$  &  $0.00490769$& $0.00755095$ & $2.271\times 10^{-6}$ & $2.218\times 10^{-6}$& 2.36 \%    \\ \cline{3-9} 
                  &                            & 0.3 & $0.011477$ &$0.0164978$ &$0.0254236$ & $6.563\times 10^{-6}$ & $6.407\times 10^{-6}$ & $2.4 \%$     \\ \cline{3-9} 
             \cline{2-9} 
                  & \textbf{0.5}    &0.2              & $0.00161291$ & $0.0042458$   & $0.0130777$ &$0.00007432$   &$0.00005560$ & $ 28.81\%$ \\ \hline\hline
                  
  \multirow{4}{*}{$\frac{1}{4}$} & \multirow{2}{*}{\textbf{0.9}} & 0.2  &  $0.00425941$ &$0.00502275$ & $0.00617356$ & $9.961\times10^{-8}$& $ 1.023\times10^{-7}$ &$2.66 \%$  \\ \cline{3-9} 
                  &                            & 0.3 & $0.0151665$ &$0.0169359$ &$0.0208834$ &$1.601\times10^{-7}$ &$1.631\times10^{-7}$ & $1.85 \%$  \\ \cline{3-9} 
             \cline{2-9} 
                  & \textbf{0.5}    &0.2              & $0.00339968$ & $0.00496717$   & $0.00743883$ &$2.12718\times10^{-6}$   &$ 1.76781\times10^{-6}$ & $18.44 \%$ \\ \hline\hline

\hline
\end{tabular}
\caption{\footnotesize Comparison of $A_{1}$ and $A_{2}$ for the EMPYM-AdS black hole using entanglement entropy.
}
\label{Table3}
\end{table}
\end{center}

Based on  Tab.\ref{Table3}, we can see that, as the pressure approaches the critical one, the relative error which translates the disagreement between Maxwell's areas decreases.
We can claim that the first order phase transition of the holographic entanglement entropy obeys to Maxwell’s equal area law just near the critical point and within our numerical accuracy.

The next obvious step in our investigation is  the check of the  critical exponent  of the second order phase transition by analyzing
the slope of the relation  between $log|T-T_c|$ and $log|\Delta S_{A}-\Delta S_{A_c}|$, where 
$\Delta S_{A_c}$ is the critical entanglement entropy found numerically by an equation $T(\Delta S_{A})=T_c$ .
We also introduce the definition of an analog to heat capacity by writing
\begin{equation}
\mathcal{C}_Q=\left.T\frac{\partial (\Delta S_A)}{\partial T}\right|_Q.
\end{equation}

\begin{center}
\begin{figure}[!ht]
\begin{tabbing}
\hspace{9cm}\=\kill
\includegraphics[scale=.78]{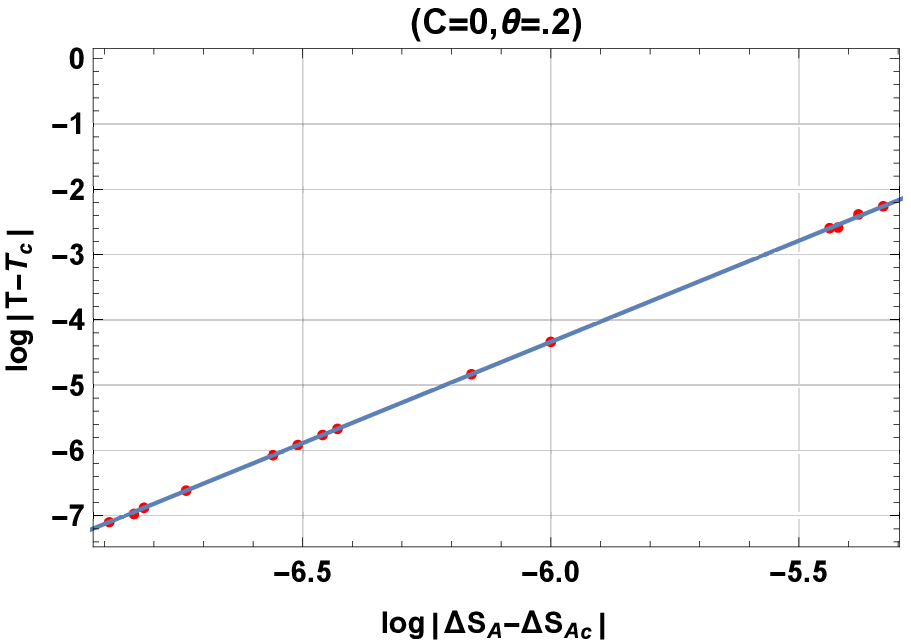}\>\includegraphics[scale=.78]{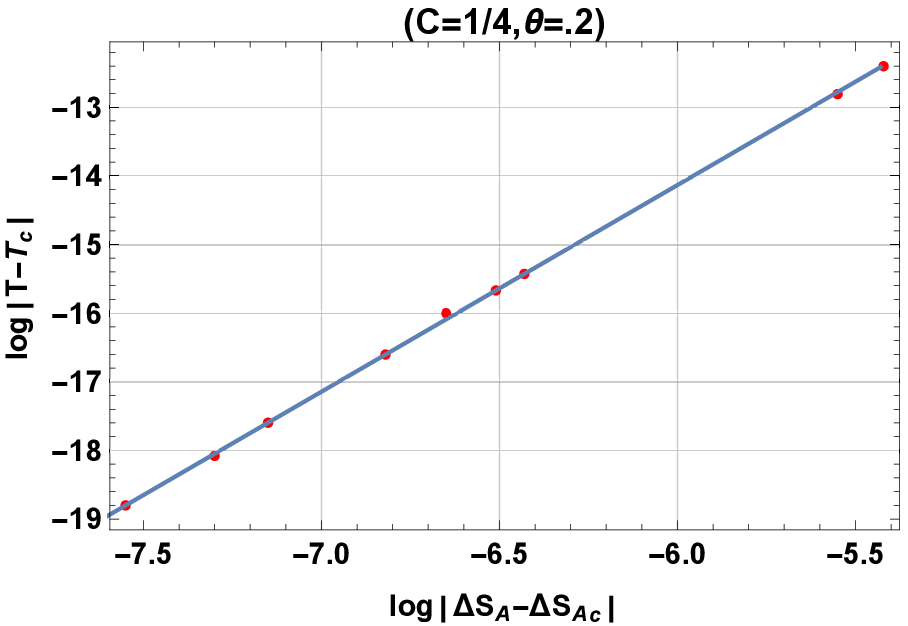} \\
\end{tabbing}
\vspace*{-.2cm} \caption{\footnotesize The relation between $log|T-T_c|$ and $log|\Delta S_{A}-\Delta S_{A_c}|$ for different $C$. {\bf(Left)}   $C=0$. {\bf (Right)}    $C\neq 0$. }\label{fig4}
\end{figure}
\end{center}

Taking $\theta_0=0.2$ and   for different  charge $C$ , we plot the relationship between $log|T-T_c|$   and $log|\Delta S_{A}-\Delta S_{A_c}|$  in Fig.\ref{fig4}, and the analytical relation can be fitted as
\begin{equation}\label{ord2EE}
log|T-T_c|=\left\{\begin{array}{cc}14.2558+ {\bf 3.09904}\ log|\Delta S_{A}-\Delta S_{A_c}|& (C=0) \\
2.9293+ {\bf 3.031061}\ log|\Delta S_{A}-\Delta S_{A_c}|& (C=\frac{1}{4}) \\
\end{array}
\right.
\end{equation}

The slop of Eq.\eqref{ord2EE}   is around  $3$ indicating that the critical exponent is $-2/3$ in total concordance with 
  that in Eq.\eqref{ord2}, therefore the critical exponent for second-order phase transition of the holographic entanglement entropy agrees with that of the thermal entropy in the canonical ensemble \cite{Zhang:2014eap,bbb} .

\begin{center}
\begin{figure}[!ht]
\begin{tabbing}
\hspace{9cm}\=\kill
\includegraphics[scale=.78]{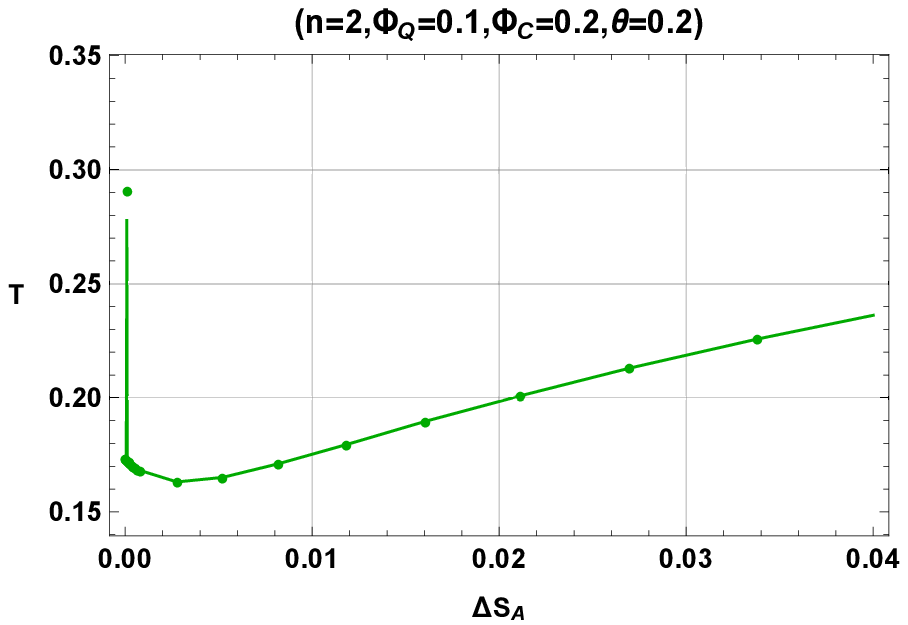}\>\includegraphics[scale=.78]{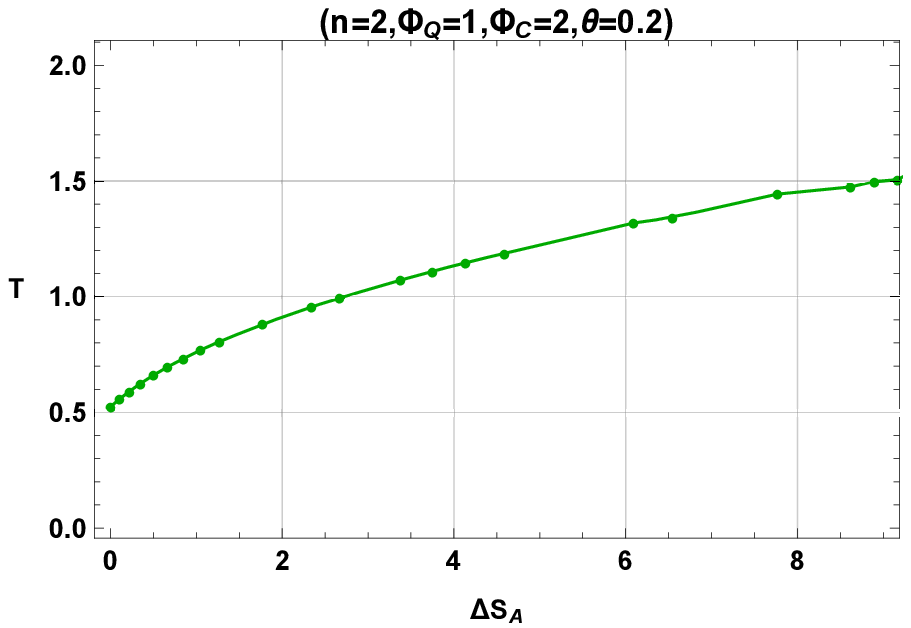} \\
\end{tabbing}
\vspace*{-.2cm} \caption{\footnotesize The relation between $log|T-T_c|$ and $log|\Delta S_{A}-\Delta S_{A_c}|$ for different $C$. {\bf(Left)}   $(\Phi_c=0.1,\Phi_Q=0.2)$. {\bf (Right)}    $(\Phi_c=1,\Phi_Q=2)$. }\label{figcan2}
\end{figure}
\end{center}

Now, we turn our attention to the grand canonical ensemble, we adopt the same analysis and the chosen  values of the previous subsection, by writing  the Eq.\eqref{fr} 
and Eq.\eqref{sysEE} as a function of the potentials $\Phi_Q$ and $\Phi_c$, taking the same boundary conditions Eq.\eqref{bc}, we perform the numerical calculations used in the plot of Hawking temperature as a function of the holographic entanglement entropy with fixed potentials in Fig.\ref{figcan2}.

Comparing Fig.\ref{figcan} with Fig.\ref{figcan2}, one may find that the thermal picture shares the same behavior of the holographic entanglement entropy, we can also observe the same minimum value of the temperature $T_{min}=0.155125$. Then the holographic framework reproduces the same attitude of the $T-S$ diagram in the grand canonical ensemble.

Now, after showing that the holographic entanglement entropy shears  the same phase picture as that of the thermal entropy for grand canonical and just near the critical point for the canonical ensemble since the  relative
disagreement between Maxwell's areas can become significantly large at low pressure.
 We attempt in the next section to explore whether the two-point correlation function has the similar behavior as that of the entanglement entropy.

\subsection{Two point   correlation function}

\begin{center}
\begin{figure}[!ht]
\begin{tabbing}
\hspace{9cm}\=\kill
\includegraphics[scale=.78]{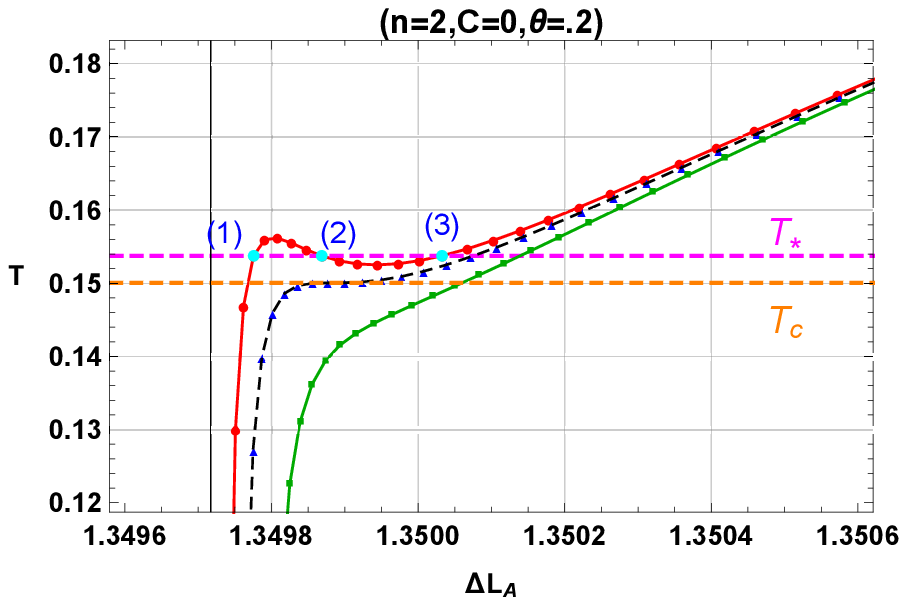}\> \includegraphics[scale=.78]{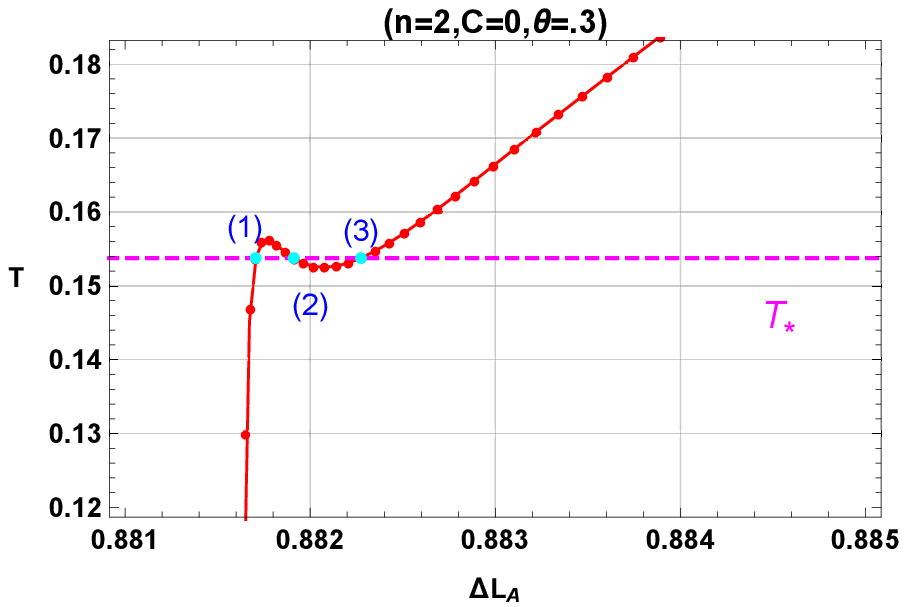} \\
\includegraphics[scale=.78]{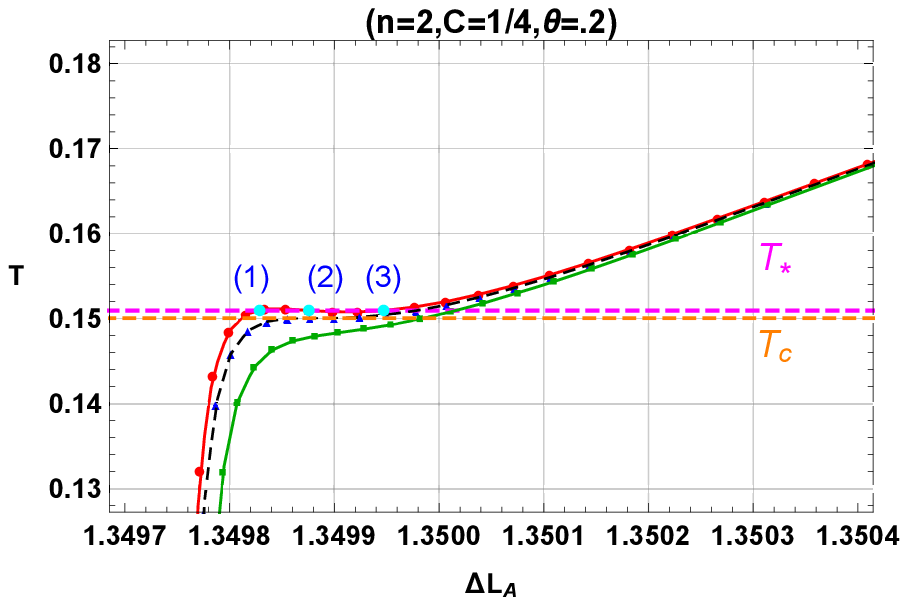}\> \includegraphics[scale=.78]{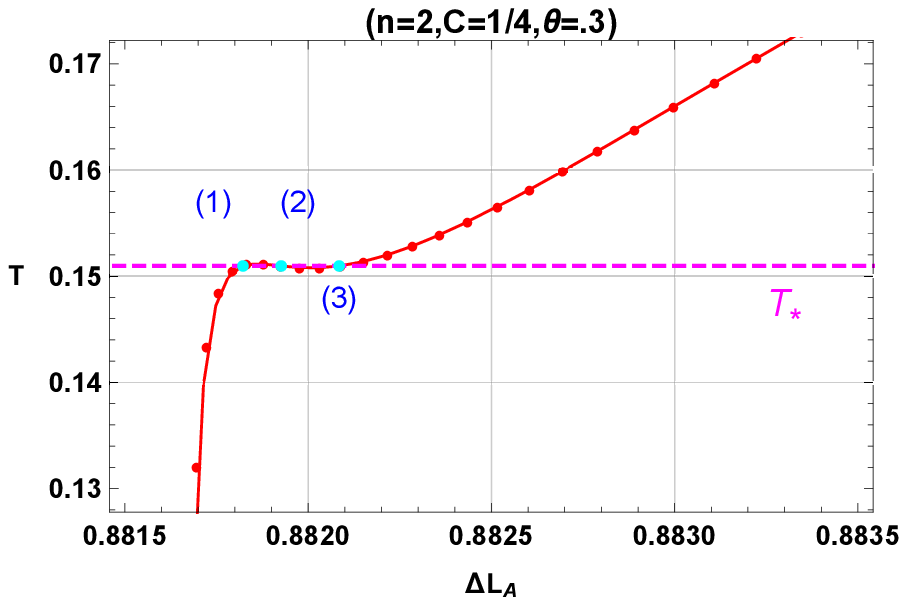} \\
\end{tabbing}
\vspace*{-.2cm} \caption{\footnotesize Plot of isocharges on the $(T,\Delta L_{A})$-plan, for $C = 0$ ({\bf left}), and  $C\neq 0$ ({\bf right}).
 For all panels: the values of the charge are $Q=0.9 Q_c$ (red), $Q=Q_{c}$ (dashed black) and $Q=2 Q_c$ (green).}\label{fig5}
\end{figure}
\end{center}

According to the Anti-de-Sitter/Conformal fields theory correspondence,
the time two-point correlation function can be written under the saddle-point assumption and in the large limit of $\Delta$ as \cite{Balasubramanian61}
\begin{equation}
\langle {\cal{O}} (t_0,x_i) {\cal{O}}(t_0, x_j)\rangle  \approx
e^{-\Delta {L}} ,\label{llll}
 \end{equation}
where $\Delta$ is the conformal dimension of the scalar operator $\mathcal{O}$ in the dual field theory, the quantity $L$ stands for the length of the bulk geodesic between the 
 points $(t_0, x_i)$ and $(t_0, x_j)$ on the AdS boundary.
Taking into account the symmetry of the considered black hole spacetime,
  we can simply  $x_i=\theta$ with the boundary $\theta_0$ and employ it to parameterize the trajectory. In this case, the proper length can be expressed as
\begin{eqnarray}
L=\int_0 ^{\theta_0}\mathcal{L}(r(\theta),\theta) d\theta,~~\mathcal{L}=\sqrt{\frac{(r^{\prime}(\theta))^2}{f(r(\theta))}+r(\theta)^2},\ \ \text{ where }\ \ \ r^{\prime}=dr/ d\theta
 \end{eqnarray}
  Treating  $\mathcal{L}$ as Lagrangian
 and $\theta$ as time, one can write the equation of motion for $r(\theta)$ as
\begin{eqnarray}
r'(\theta )^2 f'(r(\theta))-2 f(r(\theta)) r''(\theta )+2 r(\theta ) f(r(\theta ))^2=0.
\end{eqnarray}
Recalling the  boundary conditions of Eq.\eqref{bc}  we attempt to solve this equation by  choosing the same background of the previous section, in other words the same values of the parameter   $\theta_0$ with the  same UV cutoff values in the dual field theory.
The regularized two-point correlation function is labeled as $\Delta L_A=L-L_0$, where $L_0$ denotes the geodesic length in pure AdS under the same boundary region.  
In Fig.\ref{fig5} we have depicted the behavior of the temperature $T$ in function of $\Delta L_A$, the all plots show the Van der Waals-like phase transition as in the case of the thermal and the holographic entanglement entropy portrait.

As in  the case of the holographic  entanglement entropy,
the relevant calculated results are listed in Tab.\ref{Table4} which are the   $\theta_0$ values, the   $\Delta \mathcal{L}_{A_{1,2,3}}$ and the areas  $A_1$, and   $A_2$.
\begin{center}
\begin{table}[!ht]
\centering\small
\begin{tabular}{|l|l|l|l|l|l|l|l|l|}
\hline
          $C$        &         $\frac{Q}{Q_c}$                   &$\theta_{0}$ & $\Delta L_{A}^{(1)}$ &  $\Delta L_{A}^{(2)}$ & $\Delta L_{A}^{(3)}$ & $A_{1}$ &  $A_{2}$& relative error  \\ \hline\hline
          \multirow{4}{*}{$0$} & \multirow{2}{*}{\textbf{0.9}} & 0.2  &  $1.34978$ & $1.34987$  &$ 1.35003$ & $0.000287$ & $0.000300$ & $4.42 \% $   \\ \cline{3-9} 
                  &                             &$0.3$& $0.881705$ &$0.881911$ & $0.882274$ & $0.000197$ & $0.000208$  & $5.43 \% $    \\ \cline{3-9} 
             \cline{2-9} 
                  & \textbf{0.4}    &0.2              & $1.34967$ & $1.34983$   & $1.35043$ &$0.001431$   &$0.001812$ & $23.49 \%$ \\ \hline\hline
                  
  \multirow{4}{*}{$\frac{1}{4}$} & \multirow{2}{*}{\textbf{0.9}} & 0.2    &$1.34983$ & $1.34988$ &  $1.34995$ & $0.000142$ & $0.000138$  & $2.85 \%$  \\ \cline{3-9} 
                  &                            & 0.3 & $0.881822$ & $0.881926$  & $0.882085$& $0.000209$& $0.000216$  & $3.29 \%$ \\ \cline{3-9} 
             \cline{2-9} 
                  & \textbf{0.4}    &0.2              & $1.34978$ & $1.34987$   & $1.35003$ &$0.0003907$   &$0.0004653$ & $17.42 \%$ \\ \hline\hline

\hline
\end{tabular}
\caption{\footnotesize Comparison of $A_{1}$ and $A_{2}$ for the EMYM-AdS black hole using two point correlation function.
}
\label{Table4}
\end{table}
\end{center}

\newpage
The results of the Tab.\ref{Table4} tell us that under our numerical accuracy and just near the critical point 
 Maxwell's equal area law still verified implying  $A_1$   and $A_2$ are equal in the proximity to the critical pressure.
 These remarks consolidate the behavior of all panels of Fig.\ref{fig5}. At this point,
one can conclude that like the entanglement entropy, the two-point correlation function also exhibits apparently a first order phase transition as that of the thermal entropy.  However, exploring broader ranges of the pressure
has revealed the fact that
the equal area law does not also hold.

\begin{center}
\begin{figure}[!ht]
\begin{tabbing}
\hspace{9cm}\=\kill
\includegraphics[scale=.78]{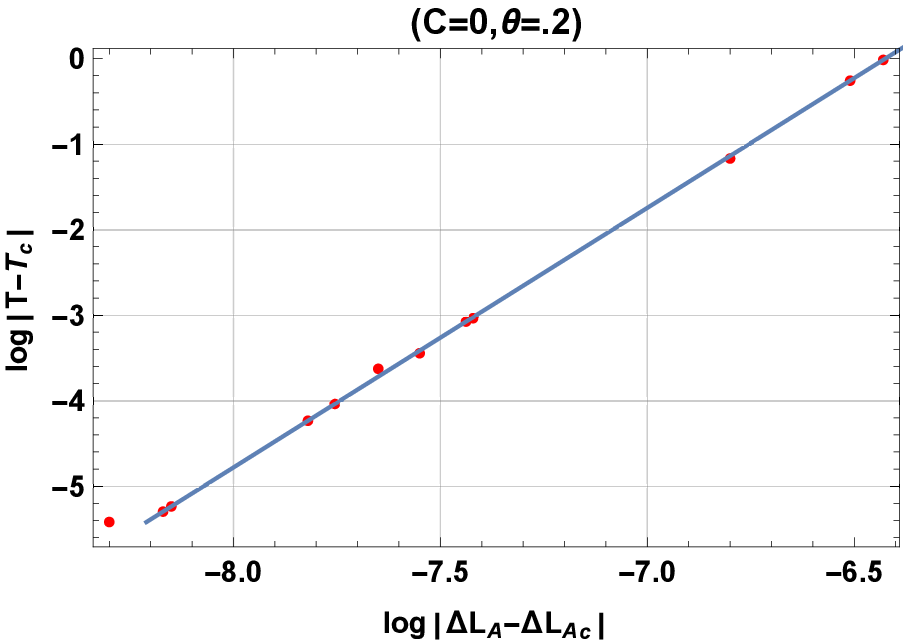}\> \includegraphics[scale=.78]{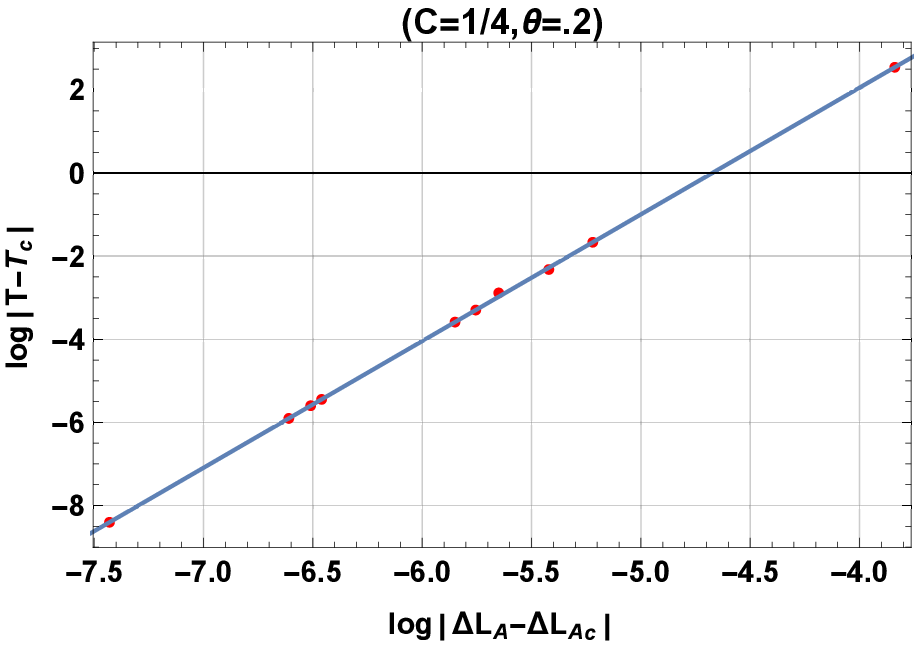} \\
\end{tabbing}
\vspace*{-.2cm} \caption{\footnotesize The relation between $log|T-T_c|$ and $log|\Delta L_{A}-\Delta L_{A_c}|$ for different charge $C$. }\label{fig6}
\end{figure}
\end{center}

For the second phase transition, we will be interested in the quantities   $log|T-T_c|$ and $log|\Delta L_{A}-\Delta L_{A_c}|$ in which $\Delta L_{A_c}$ is  the obtained numerically by the equation $T(\Delta L_{A})=T_c$. The relations between the logarithm of $|T-T_c|$ and $|\Delta L_{A}-\Delta L_{A_c}|$  are shown 
in Fig.\ref{fig6} 

The straight blue line in each panel of Fig.\ref{fig6} is fitted  following the linear equations
\begin{equation}
log|T-T_c|=\left\{\begin{array}{cc}19.056+ {\bf 3.03594}\ log|\Delta L_{A}-\Delta L_{A_c}|& (C=0) \\
14.25487+ {\bf 3.04967}\ log|\Delta L_{A}-\Delta L_{A_c}|& (C=\frac{1}{4}) \\
\end{array}\right.
\end{equation}

Again, we found a slope around 3,
then the critical exponent of the specific heat capacity is consistent with that of the mean field
theory of the Van der Waals as in the thermal and entanglement entropy portraits  \cite{Zhang:2014eap,bbb}. Therefore, we conclude that the two-point correlation function of the   Anti-de-Sitter-Maxwell-Yang-Mills black hole exists a second order phase transition at the critical temperature $T_c$.

\begin{center}
\begin{figure}[!ht]
\begin{tabbing}
\hspace{9cm}\=\kill
\includegraphics[scale=.78]{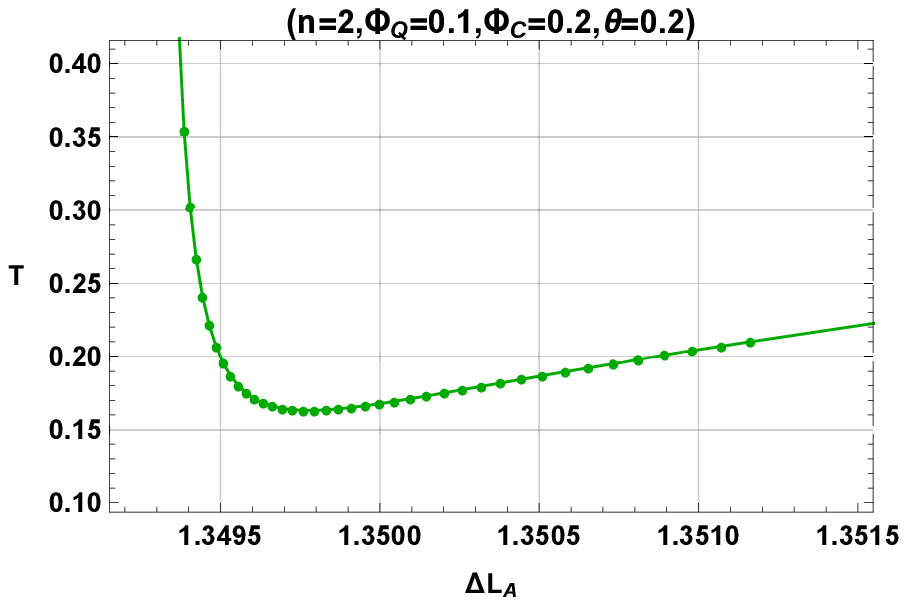}\>\includegraphics[scale=.78]{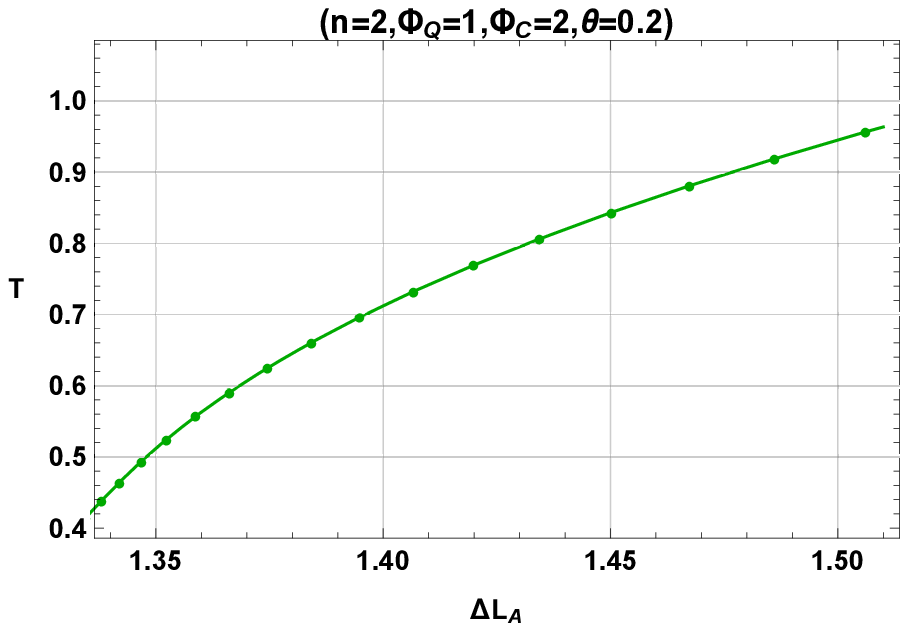} \\
\end{tabbing}
\vspace*{-.2cm} \caption{\footnotesize The relation between $log|T-T_c|$ and $log|\Delta S_{A}-\Delta S_{A_c}|$ for different $C$. {\bf(Left)}   $(\Phi_c=0.1,\Phi_Q=0.2)$. {\bf (Right)}    $(\Phi_c=1,\Phi_Q=2)$. }\label{figcan3}
\end{figure}
\end{center}

For the grand canonical ensemble, we also plot the temperature $T$ in function of $\Delta L_A$ in the Fig.\ref{figcan3}, from which we can see that thermodynamical behavior is held as the thermal and the holographic entanglement entropy frameworks.  
 
At this level  we  remark radical rupture appears when we change the thermodynamical ensemble (canonical/grand canonical). The complete comprehension  of such different behaviors is not yet completely understood. We believe  that it's typical of such system \cite{anaNPB}.

\section{Conclusion}
In this work We have investigated   the  phase transition of  Anti-de-Sitter  black hole  
in the Einstein-Maxwell-Yang-Mills gravity
  with considering the canonical and the grand canonical ensemble.
We first studied the phase structure of the thermal entropy in the $(T,S)$- plane for fixed charges and found that the phase structure agrees with the study made in \cite{Zhang:2014eap} when the electrodynamics is linear. The authors
consider the thermodynamics of such black hole in the $(P,V)$-plane notably the critical behavior and the analogy with the  Van der Waals gas. We also have shown that this behavior disappears in the grand canonical ensemble where the potentials $\Phi_Q$ and $\Phi_c$ are kept fixed. 

After, we  found  that this phase structure of the EMYM-AdS black hole can be probed by the two-point correlation function and  holographic entanglement entropy in each thermodynamical ensemble, 
which reproduce the same thermodynamical behavior of the thermal portrait just 
for a  ranges of the pressure near the critical one where the equal area law  hold within
 our numerical accuracy, for broader ranges the disagreement between Maxwell's areas becomes significant. 
These remarks remain open questions while this approach provides a new step in our understanding of the black hole phase structure from the point of view of holography. Considering the high dimensional solutions or additional hairs by adding Yang-Mills fields and taking into account their confinement 
can be the object of a totally future publication.


\end{document}